\documentclass[12pt]{iopart}

\usepackage{iopams} 
\usepackage{graphicx}

\begin{document}

\title {DDFT calibration and investigation of an anisotropic phase-field crystal model}
\author {Muhammad Ajmal Choudhary$^1$, Daming Li$^1$, Heike Emmerich$^1$ and Hartmut L\"owen$^2$}
\address{$^1$ Lehrstuhl f\"ur Material- und Prozesssimulation, Universit\"at Bayreuth, 
D-95440 Bayreuth, Germany}
\address{$^2$ Institut f\"ur Theoretische Physik II: Weiche Materie,
Heinrich-Heine-Universit\"at D\"usseldorf, 
D-40225 D\"usseldorf, Germany}
\ead{ajmal.choudhary@uni-bayreuth.de}

\begin{abstract}

The anisotropic phase-field crystal model recently proposed and 
used by Prieler et al. [J. Phys.: Condens. Matter {\bf 21}, 464110 (2009)] 
is derived from microscopic density functional theory for anisotropic particles with fixed orientation. Further its morphology diagram is explored.
In particular we investigated the influence of anisotropy and undercooling on the process of nucleation and microstructure formation from atomic to the microscale. 
To that end numerical simulations were performed varying those dimensionless parameters which represent anisotropy and undercooling in our anisotropic phase-field crystal 
(APFC) model. The results from these numerical simulations are summarized in terms of a morphology diagram of the stable state phase. These stable phases are also 
investigated with respect to their kinetics and characteristic morphological features.

\end{abstract}

\maketitle

\section{\label{sec:1} INTRODUCTION}

Early in the 90th Swift and Hohenberg formulated an amplitude approach to 
describe systems, where the stable states are periodic, 
as e.g.~the case for Rayleigh-B\'enard convection~\cite{swift}.
More recently this idea has been taken up by the materials science
community to model crystals at the atomic scale.
Elder et al.~proposed a functional for a scalar dimensionless field $\phi$ of form 
\begin{equation}\label{energyref}
F=\int\limits_V \left( \frac{1}{2}\phi
\left[(q_0^2+\nabla^2)^2-\varepsilon \right]\phi + \frac{1}{4}\phi^4 \right)\,dr \,,
\end{equation}
with two phenomenological parameters $q_0$ and $\varepsilon$
and a corresponding dynamical equation
\begin{eqnarray}\label{eq-cons}
\frac{\partial \phi}{\partial t}& = & \nabla^2 \frac{\delta F}{\delta \phi}
\end{eqnarray}
for this purpose \cite{elder02}. 
Since its introduction, this phase-field crystal (PFC) 
method \cite{elder02,karma20,karma21,karma22,karma23} 
has emerged as a computationally efficient alternative 
to molecular dynamics (MD) simulations for problems where the atomic and the continuum 
scale are tightly coupled. The reason is that it 
operates for atomic length scales and diffusive time scales.
Thus for a simple application such as diffusion in gold or copper
it runs $10^6$-$10^8$ times faster than the corresponding
MD calculation \cite{kenotes}.
In that sense it provides from point of view of
multiscale materials modeling an interesting link between the traditional phase-field 
method and MD. Moreover, a connection between classical density functional
theory of freezing and phase-field crystal modeling could be 
identified in \cite{karma21,Sven_Backofen}.  Thereby a second theoretical
foundation besides the Swift-Hohenberg amplitude equation
approach could be established.  Essentially it motivates
the application of PFC models also for spatially non-uniform
{\it non-periodic} states.

Recently the phase-field crystal method 
has been applied to a variety of different growth phenomena. One of its interesting
features is that other than the phase-field method,
in which elasticity explicitly needs to be integrated in the functional 
to be taken into account \cite{03Emm3},
it includes elastic effects inherently.  Thus it allows to simulate for
example features of crack propagation \cite{karma21} and plasticity
\cite{karma20,japaner} from the atomic to the micro-scale. To model the elastic
behavior of different kinds of materials, the parameters of the phase-field
crystal model equation can be adjusted to match the elastic moduli of a given
experimental system. However, in its most simplistic form, in which it is a
reformulation of the Swift-Hohenberg equation \cite{swift} with a conserved
dynamics as introduced by Elder {\it et al.} \cite{elder02,karma20} the
Poisson ratios which can be modeled, are restricted to $1/3$ (in the one mode
approximation). Moreover, since in the simplistic PFC model the
material is defined by only three parameters, it is restricted with respect to
the crystal lattice structures which it can describe as well. These are
triangular symmetries in two dimensions and $BCC$ symmetry in three dimensions
\cite{karma,Tapio_2010}. Another crystal symmetry applying to protein crystals in a
membrane could be obtained by including higher order correlation functions
\cite{jenspaul}.  Moreover, liquid crystals have been simulated by
combining the original phase-field crystal equation with an orientational
field \cite{Loewen_JPCM,kenliquidcrystals,Wittkowski}.

In \cite{roberttobe} we followed the above direction to extend the phase-field
to apply to a larger class of condensed matter systems taking
a different route:  We derived a generalized 
PFC model for isotropic as well as anisotropic crystal
lattice systems of arbitrary Poisson ratios
as well as condensed matter systems built-up from 
non-spherical units such as for example anisotropic colloids and liquid crystals.
To this end we extended the simplest
PFC model proposed by Elder \cite{elder02},
to a conservative, anisotropic Langevin
equation and applied it to the heterogeneous
nucleation of colloids at a wall.

Whereas our previous work \cite{roberttobe} was devoted to a
mere introduction of our model and its application to
heterogenous nucleation at a wall, here we show for the first
time how its parameters can be derived from dynamical density
functional theory (DDFT) (see \sref{hartmut}).  Further
we report for the first time in detail on its morphology
diagram.  To do so, we proceed as follows:  
First we give a thorough derivation of
our anisotropic phase-field model based on the DDFT in \sref{hartmut}.
We then - after a brief summary of the model in \sref{hartmut} -
study the influence of anisotropy and undercooling on the morphology
of the final states.  We analyze these
morphologies and summarize our results in terms of
a morphology diagram in \sref{sec:4}.
Finally we conclude with a summary 
and an outlook of our study in \sref{sec:5}.

\section{\label{hartmut} ANISOTROPIC PHASE-FIELD CRYSTAL MODEL}

\subsection{\label{sec:2a}The model}

By now phase-field crystal (PFC) modeling is widely used to predict crystal nucleation growth and to model microstructural pattern formation during different physical 
phenomena such as solidification. As usual, the PFC model used in this studies is based on a free energy fuctional $F[\phi(r,t)]$ of phase-field $\phi (r,t)$ and a 
dynamical equation which represents the time evolution of the phase-field. In this PFC model, periodic nature of a crystal lattice is incorporated by using a free energy
 functional which is minimized by periodic density field. The equation of motion used in this model was introduced in \cite{elder02} for the case of simplest phase-field 
crystal (SPFC) model, and is given by
\begin{equation}\label{eq:motion}
\rho \frac{\partial \phi}{\partial t} = \Delta \left[ \{ \left( q_0^2 + \Delta \right)^2 - \epsilon \} \phi +\phi^3 \right], 
\end{equation}
\noindent here, $q_0$ and $\epsilon$ are constants. In order to simplify the model, a dimensionless parameter $\tau$ is introduced which is defined as
\begin{equation}\label{eq:tau1}
\tau = - \left( q_0^2 - \epsilon \right). 
\end{equation}
\noindent In our model $\tau$ represents undercooling in the same manner as $r$ is defined in \cite{elder04}, therefore, $\tau$ can be written as
\begin{equation}\label{eq:tau2}
\tau \propto \Delta T. 
\end{equation}
The anisotropic version of phase-field crystal model originally introduced in \cite{emmerich1} is used in this study. This PFC model is basically an extension of the 
SPFC model which is derived in \cite{Boettinger1}. The APFC model is capable of simulating isotropic and anisotropic crystal lattice system of any arbitrary poisson 
ratio as well as condensed matter systems such as colloids and liquid crystals. The free energy functional used in this model is given by
\begin{eqnarray}\label{eq:APFC}
F = \int\limits_V\ \Big( \frac{1}{2}\phi \Big[ -\tau + a_{ij} \frac{\partial^2}{\partial x_i \partial x_j}
+ b_{ijkl}\frac{\partial^4}{\partial x_i \partial x_j\partial x_k \partial x_l}\Big]\phi 
+ \frac{1}{4} c \phi^4 \Big) dr,
\end{eqnarray}
\noindent where $a_{ij}$ is a symmetric matrix and $b_{ijkl}$ is a fourth rank tensor with the symmetry of an elastic tensor: $i\leftrightarrow j, k\leftrightarrow l, 
(i,j)\leftrightarrow (k,l)$. From the free energy functional defined in \Eref{eq:APFC}, the corresponding Langevin differential equation of motion for anisotropic lattice 
system can be written as follows:
\begin{eqnarray} \label{eq:langevin}
\rho \frac{\partial \phi}{\partial t} = \Delta \Big( \Big[ -\tau + a_{ij} \frac{\partial^2}{\partial x_i \partial x_j}
+ b_{ijkl}\frac{\partial^4}{\partial x_i \partial x_j\partial x_k \partial x_l}\Big]\phi
+ c \phi^3 \Big).
\end{eqnarray}
\subsection{\label{sec:2b}Derivation of the anisotropic phase-field crystal model from dynamical density functional theory}

The coefficients occuring in the anisotropic phase-field crystal model proposed in Ref.\ \cite{roberttobe} can be 
derived from microscopic density functional 
theory \cite{Evans,Singh:91,Loewen:94,Loewen:02}. Here we follow a similar line as proposed
recently by van Teeffelen et al \cite{Sven_Backofen} for radially symmetric interactions.
We generalize this route here to anisotropic interactions.

We assume that the anisotropic colloids  are completely aligned  in space. Cartesian 
coordinates ${\bf r} = (x_1,x_2,...,x_d)$ will be used in the following, $d$ denoting the spatial dimension.
The interaction pair potential between two aligned particles is $u({\bf r})$\cite{footnote}.
The latter function is anisotropic, in general, i.e.\ it does not only depend on 
$| {\bf r} |$. Other examples for these anisotropic interactions with 
fixed orientations are oriented hard spherocylinders \cite{Bohle} and charged rods \cite{JCP_2004,Kirchhoff}, 
anisotropic Gaussian potentials \cite{Saija}, board-like colloidal particles \cite{Vroege},
colloidal molecules \cite{Pine},
as well as patchy colloids \cite{Sciortino} 
and proteins \cite{Allahyarov,Allahyarov2}. Henceforth {\it inversion} symmetry is assumed
\begin{equation}\label{eq:inversion}
u(-{\bf r}) =  u({\bf r})
\end{equation}
Dynamical density functional theory for anisotropic situations \cite{RexWensink}
is now generalized from the isotropic case as follows. The dynamical evolution 
of the time-dependent one-particle density field $\rho({\bf r},t)$ is:
\begin{eqnarray}\label{eq:ddft}
  \dot\rho({\bf r},t)
  =(k_BT)^{-1} \nabla \cdot \left[ {\mathcal{D}} \rho({\bf r},t) \nabla
\frac{\delta F\left[\rho({\bf r},t)\right]}{\delta \rho({\bf r},t)}\right]\,.
\end{eqnarray}
Here $k_BT$ is the thermal energy, and 
$\nabla = ( \partial/\partial x_1,\partial/\partial x_2,...,\partial/\partial x_d)$
is the $d$-dimensional gradient.
${\mathcal{D}}=diag(D_1,D_2,...,D_d)$ denotes the diagonalized  diffusion tensor with the 
anisotropic short-time translational diffusivities of the anisotropic particle. 
For a given (hydrodynamic) shape of the particle, explicit expressions for $D_i$
are available \cite{delaTorre,LoewenBD}. Furthermore, in \Eref{eq:ddft}, $F\left[\rho({\bf r},t)\right]$ is the equilibrium density 
functional which can be split as
\begin{equation}\label{eq:ftot}
F\left[\rho({\bf r})\right]= F_{\rm{id}}\left[\rho({\bf r})\right]
+ F_{\rm{ex}}\left[\rho({\bf r})\right]+F_{\rm{ext}}\left[\rho({\bf
    r})\right]\,.
\end{equation}
where
\begin{equation}\label{eq:fid}
F_{\rm id}\left[\rho({\bf r})\right]= k_B T \int\,{\mathrm d} {\bf r} 
\rho({\bf r})\left\{\ln\left[\rho({\bf r})\Lambda^d\right]-1\right\}\,,
\end{equation}
with $\Lambda$ denoting the thermal de Broglie wavelength. 
The external part involves an external one-body potential $V({\bf r},t)$
and  is given by
\begin{equation}\label{eq:fext}
  F_{\rm{ext}}\left[\rho({\bf r})\right]=\int\,{\mathrm d} {\bf
    r}\rho({\bf r})V({\bf r},t)\,.
\end{equation}
Finally, the excess part $F_{\rm ex}[\rho({\bf r})]$, embodies the nontrivial
correlations between the particles and must
be further approximated. Henceforth we assume small deviations of the 
inhomogeneous density profile around a homogeneous reference density 
$\rho$. In this limit, the leading approximation for $F_{\rm{ex}}\left[\rho({\bf r})\right]$
is given by the Ramakrishnan and Yussouff~\cite{Ramakrishnan:79} expression:
\begin{eqnarray}\label{eq:fex}
F_{\rm{ex}}\left[\rho({\bf r})\right]\simeq F_{\rm{ex}}(\rho)
-\frac{k_BT}{2}\int\int{\mathrm d} {\bf r}{\mathrm d}
{\bf r}^\prime \Delta\rho({\bf r})\Delta\rho({\bf
  r}^\prime)c_0^{(2)}({\bf r}-{\bf r}^\prime;\rho)\,.
\end{eqnarray}
where $c_0^{(2)}({\bf r}-{\bf r}^\prime;\rho)$ is the anisotropic direct correlation 
function of the fluid at density $\rho$ which possesses the same symmetry as 
the underlying pair potential $u({\bf r})$. In particular, it is inversion-symmetric
\begin{equation}\label{eq:inversion_c}
c_0^{(2)}(-{\bf r}, \rho) = - c_0^{(2)}({\bf r}, \rho)
\end{equation}
Moreover, $\Delta\rho({\bf r})=\rho({\bf r}) - \rho$. In Fourier space \Eref{eq:fex}
reads 
\begin{eqnarray} \label{eq:PFC-Fexc}
\mathcal F_{\rm ex}[\rho({\bf r})]&=&F_{\rm ex}(\rho)
- \frac{k_{\rm B}T (2\pi)^d}{2} 
\int {\rm d {\bf k}} {\Delta \tilde {\rho}}({\bf k}) {\Delta {\tilde \rho}}({\bf -k})\tilde
c_0^{(2)}({\bf k},\rho)
\end{eqnarray}
with $\sim $ denoting a Fourier transform. We now expand the direct correlation function 
$c_0^{(2)}({\bf k},\rho)$ in terms of $\bf k$ around ${\bf k}=0$. (Alternatively 
 fitting procedures can be used, e.g. around the first peak of $c_0^{(2)}({\bf k},\rho)$.)
This leads to the Taylor expansion in Fourier space 
\begin{eqnarray} \label{eq:Taylor}
\tilde c_0^{(2)}({\bf k},\rho) = \hat{C}_0 + \sum_{i,j=1}^d a_{ij} k_i k_j
+ \sum_{i,j,k,l=1}^d b_{ijkl} k_i k_j k_k k_l +\dots \,.
\end{eqnarray}
corresponding to a gradient expansion in real-space. Inversion symmetry \Eref{eq:inversion_c}
enforces all odd orders to vanish. Possible additional symmetries in the shape of the 
particles will lead to corresponding restrictions on the tensorial coefficients $a_{ij}$ 
and $b_{ijkl}$ as discussed below.

Inserting this expansion into \Eref{eq:ddft}, one gets
\begin{eqnarray} \label{eq:PFC1}
\dot\rho({\bf r},t) = &\nabla& \cdot  {{\mathcal{D}}}\nabla \rho({\bf r},t)
+\nabla \cdot  {{\mathcal{D}}}\nabla \Big[ (k_BT)^{-1} V({\bf r},t)\nonumber \\
&-& ( \hat{C}_0- \sum_{i,j=1}^d a_{ij}\frac{\partial^2}{\partial x_i \partial x_j}
 +  \sum_{i,j,k,l=1}^d b_{ijkl} \frac{\partial^4}{\partial x_i \partial x_j \partial x_k \partial x_l}) \rho({\bf r},t)  \Big]\,.
\end{eqnarray}
If one further uses the constant mobility approximation, 
$\rho({\bf r},t) = \rho$ in front of the functional 
derivative in \Eref{eq:ddft} and if one approximates
\begin{eqnarray}\label{eq:fid_cpfc}
F_{\rm id}\left[\rho({\bf r})\right]\approx k_B T\rho \int\,{\mathrm d} {\bf r} 
\Big\{\frac{1}{2}\phi({\bf r},t)^2-\frac{1}{6}\phi({\bf r},t)^3
+\frac{1}{12}\phi({\bf r},t)^4-{\rm const.}\Big\}
\end{eqnarray}
with $\phi({\bf r},t)=\Delta \rho({\bf r},t)/\rho$, one arrives at: 
\begin{eqnarray}\label{eq:PFC2}
     \dot\phi({\bf r},t) &=& \rho \nabla \cdot  {{\mathcal{D}}}\nabla \Bigg[ \phi({\bf r},t) 
    - \frac{1}{2}\phi({\bf r},t)^2 +  \frac{1}{3} \phi({\bf r},t)^3 + (k_BT)^{-1}V({\bf r},t) \nonumber\\
    &-& \rho ( \hat{C}_0 -\sum_{i,j=1}^d a_{ij} \frac{\partial^2}{\partial x_i \partial x_j}
 +  \sum_{i,j,k,l=1}^d b_{ijkl} \frac{\partial^4}{\partial x_i \partial x_j \partial x_k \partial x_l}
 ) \phi({\bf r},t) \Bigg] \,.
\end{eqnarray}
This exactly reduces to the anisotropic phase-field model  of Ref.\ \cite{roberttobe} for the special case $d=2$, 
${\mathcal{D}} = D_0 {\mathcal{1}}$, and a neglected cubic term in the ideal gas functional expansion 
in \Eref{eq:fid_cpfc}. As a remark the latter was retained in other variants of the PFC model
\cite{Elder,Tapio}.

Concluding this section, the anisotropic phase-field crystal
 model as used in \cite{roberttobe} can be derived and justified from dynamical
density functional theory. 
The derivation points, however, to more realistic approximations for anisotropic 
diffusivities. Furthermore, if \Eref{eq:fid_cpfc} is used, some approximations can be avoided but
these were not found to change the results significantly for spherical 
interactions \cite{Sven_Backofen}.

\subsection{\label{sec:2c}Phenomenological symmetry considerations}

We finally present phenomenological symmetry arguments for the expansion coefficients $a_{ij}$ and $b_{ijkl}$
of the anisotropic PFC model. First we assume that the orientation of the fixed particles is set by a single unit vector ${\vec E}$ only
which is invariant under space inversion (${\vec r} \to -{\vec r}$). This is
 the case for $d=2$ and for rotationally symmetric particles in $d=3$. Then, any gradient term in the scalar free 
energy functional must involve an even number of gradients due to space inversion symmetry. Rotational symmetry 
of space then requires that only combinations of ${\vec E} \cdot {\vec \nabla}$ and ${\vec \nabla} \cdot {\vec \nabla}$
are nonvanishing in the functional. Therefore the only possibility for physically relevant gradient terms is
\begin{eqnarray}\label{eq:coeff1}
   \sum_{i,j=1}^d a_{ij} \frac{\partial^2}{\partial x_i \partial x_j} =
\lambda_1 ({\vec E} \cdot {\vec \nabla})^2 + \lambda_2 \Delta
\end{eqnarray}
and
\begin{eqnarray}\label{eq:coeff2}
     \sum_{i,j,k,l=1}^d b_{ijkl} \frac{\partial^4}{\partial x_i \partial x_j \partial x_k \partial x_l}
= \lambda_3 ({\vec E} \cdot {\vec \nabla})^4
+ \lambda_4 ({\vec E} \cdot {\vec \nabla})^2 \Delta + \lambda_5 \Delta^2
\end{eqnarray}
where $\lambda_1$, $\lambda_2$, $\lambda_3$, $\lambda_4$, and $\lambda_5$ are scalar prefactors.
This reduces the number of independent degrees of freedom in $a_{ij}$ and $b_{ijkl}$ down to 5.

In case there are different fixed vectors, say ${\vec E}$ and ${\vec B}$, there are  
corresspondingly more terms allowing for more freedom in $a_{ij}$ and $b_{ijkl}$. This 
is realized, e.g., for biaxial colloidal particles in two crossed external fields along  ${\vec E}$ and ${\vec B}$.

\section{\label{sec:3} SIMULATION PARAMETERS}

As initial condition, a square domain is defined with a sphere in the center to initialize the nucleus. Periodic boundary conditions in all directions in the square box 
are used. The values for the parameters are the same as defined in \cite{roberttobe}, namely $\rho = 1$ (which sets the time scale), 
$a_{11} = a_{22} = 2$, $b_{1111} = b_{2222} = b_{1122} =1$, $b_{1212} = 0$ and $c = 1$, respectively \cite{foot}. However, a typical set of values for $\tau$ and $s$ is 
used for each simulation since our basic objective is to study the dependence of the stable state phase on these parameters. A simple explicit numerical scheme is used to 
obtain a reasonably well approximated solution. A forward Euler scheme is used for the time derivative with a sufficiently small time step of $\Delta t=$0.00075 to ensure 
the stability of the scheme. The Laplace operator is approximated by using second order difference scheme given by

\begin{eqnarray} \label{eq:scheme}
\nabla^2 \phi = \Big(\phi _{i+1,j} + \phi _{i-1,j} + \phi _{i,j+1} 
+ \phi _{i,j-1} - 4\phi _{i,j}\Big)/ ( \Delta x)^2.
\end{eqnarray} 

For the following simulations $\Delta x$ is chosen as $\Pi/4$.  Convergence
of our results was ensured via convergence studies.  The morphologies
depicted in \fref{fig:1_anisotropy}, \fref{fig:2_undercooling} and \fref{fig:4_morphologies} show simulations of 256 times 256
numerical grid units. However, morphologies in \fref{fig:5_stripes} show simulation of 128 times 256 numerical grid units.

\section{\label{sec:4} RESULTS AND DISCUSSION}

In this section we present the simulation results obtained from our studies. These simulation results demonstrated the following issues concerning nucleation and successive
 microstructure formation: \\
  (i) The effect of undercooling on crystal growth. \\
 (ii) The dependence of anisotropy and undercooling on stable state phase.\\
(iii) The effect of anisotropy and undercooling on distance between the neighbouring stripes. 

\subsection{\label{sec:4a} Anisotropic effects}

In order to quantify the anisotropy of the material at the atomic scale, a dimensionless parameter $s$ is introduced and is given by

\begin{equation}\label{eq:shear}
s = - \frac{b_{1112}}{b_{1111}}. 
\end{equation}

\begin{figure} [htbf]
\centering
      \includegraphics[width= 8.55 cm , height = 4.3 cm]{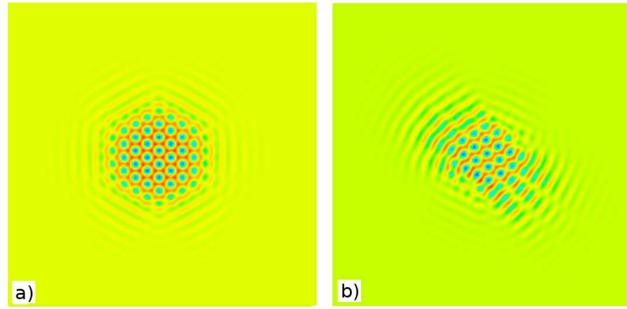}
\caption{Simulation results of crystal growth calculated for $\tau = -3/4$ and (a) $s = 0$ (isotropic case) and (b) $s = 0.125$ (anisotropic case). 
All other parameters are chosen as given in \sref{sec:3}.  As expected the isotropic case shows a symmetric
morphology.}
\label{fig:1_anisotropy}
\end{figure}

The effect of this dimensionless parameter is studied by performing numerical simulations with $s = 0$ and $s = 0.125$. The initial and boundary conditions as well as the 
values for all other simulation parameters used in these simulations as same as given in \sref{sec:3}.
The results obtained for both cases after 30,000 time steps are shown in \fref{fig:1_anisotropy}.

\subsection{\label{sec:4b} Undercooling effects}

When a liquid is supercooled just below the melting temperature the crystal starts growing and the crystal growth is directly related to the undercooling. 
Depending on the formal undercooling, which quantifies the distance from the phase-equilibrium line in the phase diagram, in the system the final state will have 
different morphologies.  These are categorised and analysed here in detail to get an overview of the state phase of the APFC model given by the variables $s$ and $\tau$.  

\begin{figure} [htbf]
\centering
      \includegraphics[width= 8.55 cm , height = 4.3 cm]{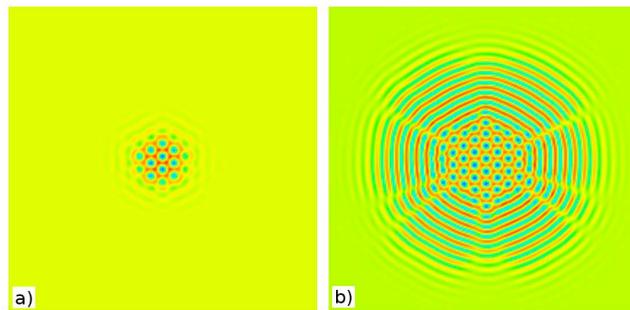}
\caption{Simulation results of crystal growth after 40,000 time steps for $s = 0$ and (a) $\tau = -0.8$ and (b) $\tau = -0.25$}
\label{fig:2_undercooling}
\end{figure}

In this section the APFC model is used to examine the rate of crystal growth from a supercooled liquid state. As explained above, $\tau$ represents the undercooling in 
our model. A number of simulations with different values of $\tau$ are performed for a specific $s$ value. We used the same initial condition i.e. single solid nucleus 
(nucleation site) for all simulations. The results showed that the rate of crystal growth increases with an increase in the value of $\tau$. As an example the simulation 
results with $s = 0$ and for two typical values of $\tau$ i.e. with $\tau = -0.25$ and $\tau = -0.8$ respectively at 40,000 time steps are depicted in 
\fref{fig:2_undercooling}. 

\subsection{\label{sec:4c} Heterogeneous nucleation and crystal growth}

\begin{figure} [htbf]
\centering
      \includegraphics[width= 8.5 cm , height = 5.8 cm]{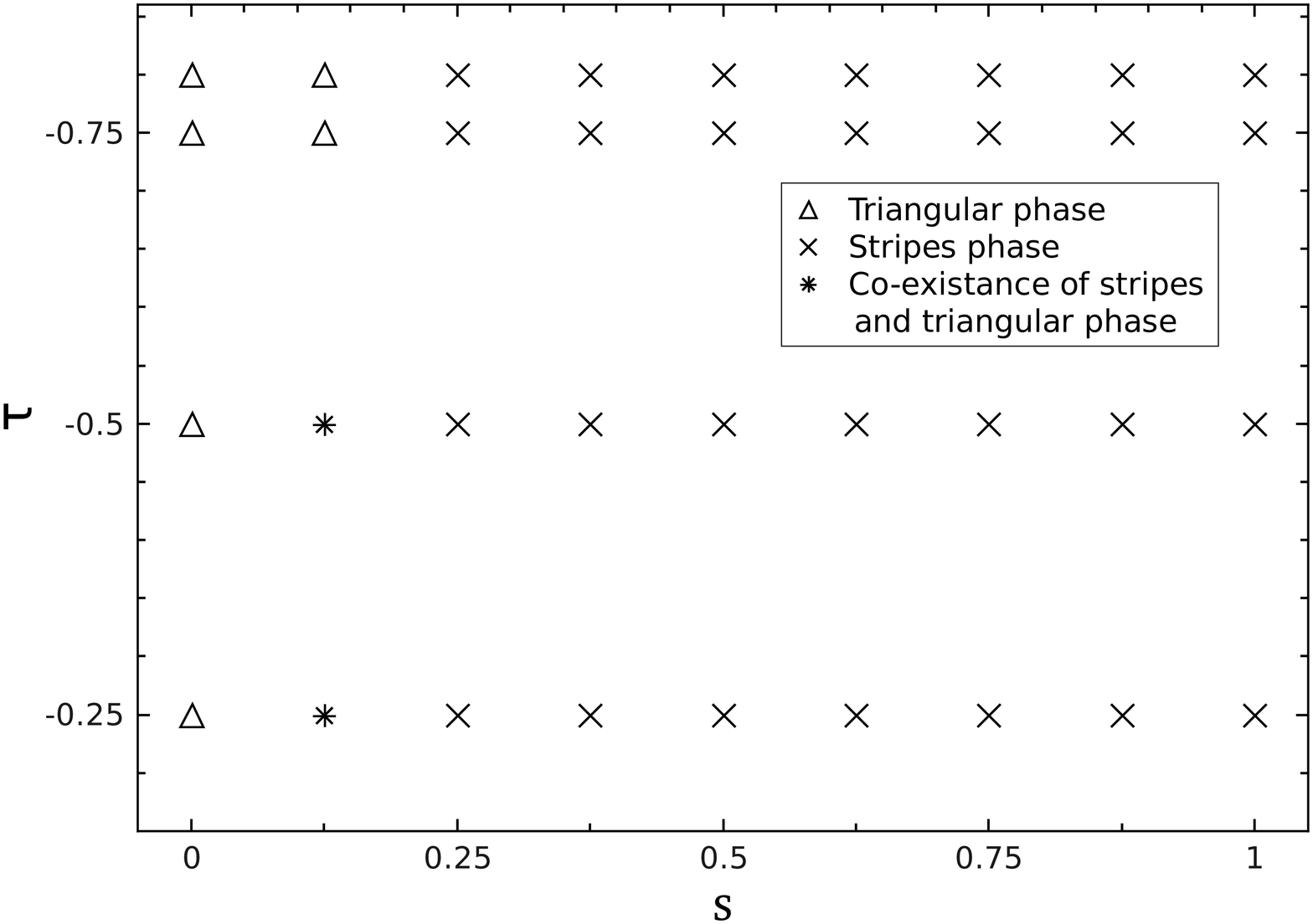}
\caption{Simulation results of  stable phases for same initial conditions and different values of $s$ and $\tau $}
\label{fig:3_different_phases}
\end{figure}

In this section we studied the dependence of a stable state phase on anisotropy and undercooling. More specifically, we show how a stable state can be composed of different 
phases such as triangular phase, stripe phase and co-existance of stripes and triangular phase, depending on the values of $s$ and $\tau$ i.e. anisotropy and undercooling 
respectively. To study these patterns we performed a number of simulations with different values of $s$ and $\tau$ in each simulation. However, the other parameters for all 
these simulations are same as described in the previous section. The results in the form of 
a diagram of stable state phase are demonstrated in \fref{fig:3_different_phases}. 

\begin{figure} [h]
\centering
      \includegraphics[width= 8.7 cm , height = 8.58 cm]{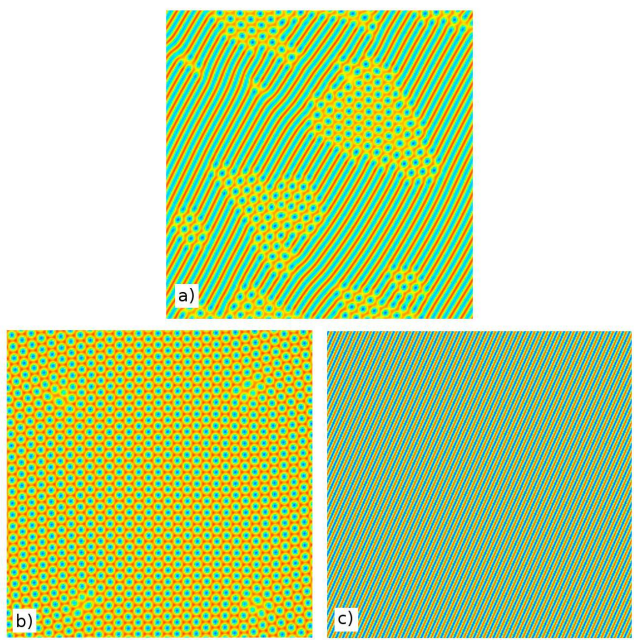}
\caption{Typical stable states phases from simulations performed for (a) $s = 0.125$ and $\tau = -0.25$, which result in a co-existence of stripes and triangular phase, 
(b) $s = 0$ and $\tau = -0.75$ which result in the triangular phase , and (c) $s = 0.75$ and $\tau = -0.25$, which results in stripes}
\label{fig:4_morphologies}
\end{figure}

These simulation results demonstrate that the stable state phase always consists of stripes if $s \geq 0.25$ irrespective of the $\tau$ value. The co-existence of 
stripes and triangular phases is found only in case of $s = 0.125$ and $\tau \geq -0.5$, while for other values of $s$ and $\tau$, the stable state consist of a 
triangular phase. Typical picture of the triangular phase, stripes, and a co-existence of stripes and triangular phase are given in \fref{fig:4_morphologies}. 

The precise shape of the grain boundary between the two solid phases depends on the system size due to \cite{nonroteffect}. 
The two-phase coexistence as such, however, is independent of the size of the system.

\subsection{\label{sec:4d} Distance between the neighbouring stripes}

\begin{figure} [h]
\centering
      \includegraphics[width= 8 cm , height = 5.5 cm]{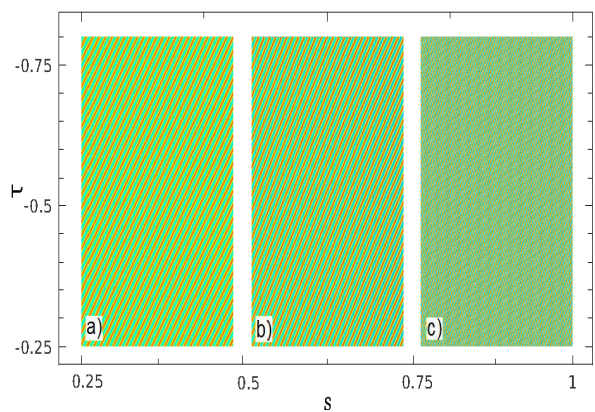}
\caption{Stripe phases for different values of $s$ and $\tau$. Resulting stripe phase for (a) $s = 0.25$, (b) $s = 0.625$ and (c) $s = 1$.}
\label{fig:5_stripes}
\end{figure}

To analyse the stripe morphology further, we studied the effect of anisotropy and undercooling on the distance between the neighbouring stripes. 
As discussed in the previous section, the stable state consists of stripes when $s \geq 0.25$ irrespective of the $\tau$ value, as shown in \fref{fig:3_different_phases}. 
It is observed that the stripe phase obtained for different values of $s$ are different from each other in terms of the spacing between the neighbouring stripes in the 
stripe phases. However, simulations with different values of $\tau$ result in similar stripe phases. Our results reveal that the spacing between neighbouring stripes 
decreases with an increase in the value of $s$.  Thus the stripe phase obtained  with $s = 1$ are much finer compared to the ones obtained with $s = 0.25$.  
Further, to investigate the effect of the undercooling $\tau$ on the stripe phase, we performed several simulations by fixing a specific value for $s$ and varying $\tau$. 
The results show that for a specific value of $s$, similar stripe morphologies are obtained with different value of $\tau$. \Fref{fig:5_stripes} demonstrates this 
interesting finding for different values of $s$ and $\tau$. 

From the above discussion, we can conclude that the distance between neighbouring stripes decreases as we increase the value of $s$. However, the undercooling has no 
significant effect on the stable stripe phase, i.e.~the stripe phases for different values of $\tau$ are similar.      

\subsection{\label{sec:4e} Analysis of the anisotropy introduced by s}

\begin{figure} [h]
\centering
      \includegraphics[width= 8 cm , height = 5.5 cm]{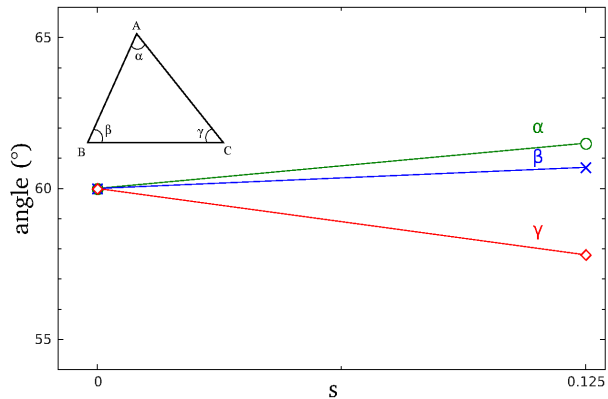}
\caption{Form of triangles  in the final triangular phase obtained for different anisotropies $s$.}
\label{fig:6_angles}
\end{figure}

In this section we discuss the influence of the anisotropic factor $s$ on the final triangular phase obtained with $0\leq s \leq 0.125$ and $\tau = -0.75$. 
More specifically, we analysed the form of triangles  in each of the final triangular phase. The form of a triangle is determined in terms of three internal angles 
and ratio of length of the longest and shortest sides of the triangle.
It is observed that the triangular phase obtained in case of $s = 0$ i.e. without any anisotropy, consist of triangles with same three angles of $60^o$ each. 
However, the triangular phase obtained with non-zero values of $s$ contains triangles with dissimilar sides. The details of the angles calculated for each case are 
shown in \fref{fig:6_angles}.
 
\begin{figure} [ht]
\centering
      \includegraphics[width= 8 cm , height = 5.5 cm]{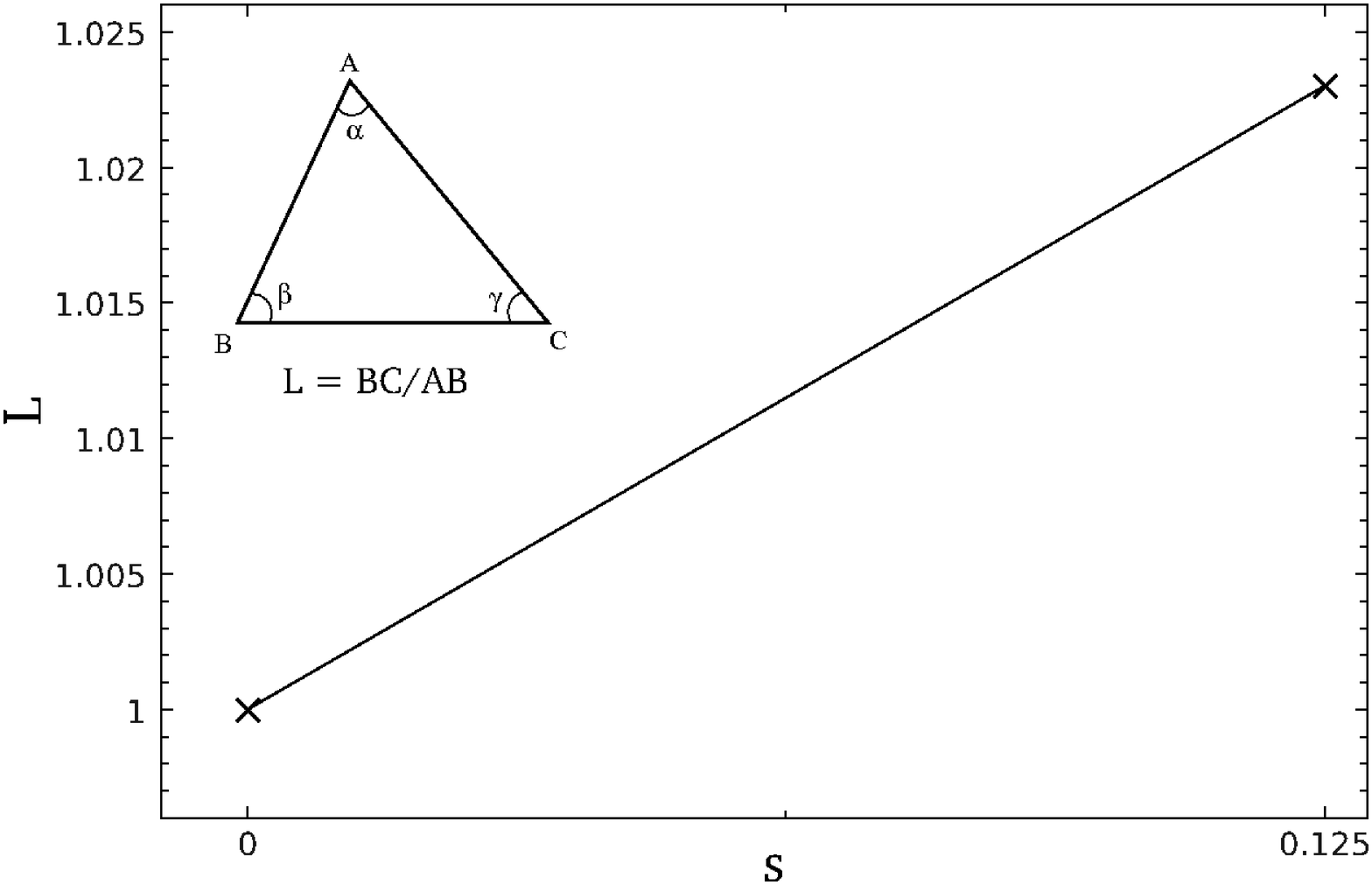}
\caption{Ratio of the length of longest and shortest sides of the triangles  in the final triangular phase versus $s$.}
\label{fig:7_length_ratio}
\end{figure}

These results demonstrate that for isotropic case i.e. $s = 0$, the final triangular phase consist of equilateral triangles. However, for anisotropic case i.e. non zero values of $s$, the final triangular
phase contain scalene triangles i.e. no two sides are similar. The ratio of length of the longest and shortest sides of the triangles, calculated for each case is presented in \fref{fig:7_length_ratio}. 
This underlines the capability of our APFC model to give rise to truly anisotropic morphologies.

\section{\label{sec:5} SUMMARY AND OUTLOOK}

In this article we presented a DDFT based derivation of the APFC model proposed by two of the authors (D.~L.~and H.~E.)
in \cite{roberttobe} previously.  Further we investigated the state phase of this model given by
variation of $\tau$ and $s$ to demonstrate its capacity to model structures
beyond those captured by the SPFC equations originally introduced
by Elder et al.~\cite{elder02}.

In particular we studied the influence of anisotropy and undercooling on final states using numerical techniques to minimize the free energy functional in our model. More specifically, a number of numerical simulations are performed by using different sets of values for our dimensionless parameters $s$ and $\tau$ which represent anisotropy and undercooling, respectively.
The results obtained from these numerical simulations are analysed. Our studies reveal that:\\
 (i) the rate of crystal growth increases with increase of $\tau$ i.e. undercooling; \\
 (ii) the stable state phase consists of a stripe phase if $s \geq 0.25$ irrespective of $\tau$. However, the stable state is a co-existence of stripes and triangular phase when $s = 0.125$ and $\tau \geq -0.5$, while for other values of $s$ and $\tau$, the stable state consists of a triangular phase;\\
(iii) for $s \geq 0.25$, the undercooling $\tau$ has no effect on the resulting stripe phases, however, the distance between the neighbouring stripes decreases with an increase of $s$.\\
(iv) Triangular crystals with cells that are neither equilateral nor rhombic are possible for our anisotropic model. 

In the future we plan to extend the approach to reactive systems \cite{gemming}
to simulate morpholgenesis in such systems from the atomic to microscale.

\section{\label{sec:6} ACKNOWLEDGMENTS}

We thank S. van Teeffelen and T. Ala-Nissila for helpful discussions. 
This work has been supported by the DFG through 
the DFG priority program SPP 1296. 
Daming Li is also supported by National Sciences Foundation 
of China (Grant No. 10701056).


\begin{thebibliography}{8}
\bibitem{footnote} By scaling, hard ellipsoids with fixed orientation are formally
    equivalent to hard-sphere systems. This is no longer true, of coarse, if the interactions
    are soft and involve an explicit energy scale.
\bibitem{foot} This choice of parameters does not correspond to a single field 
direction ${\vec E}$ but to two crossed fields $\vec E$ and $\vec B$, see II.C.
\bibitem{nonroteffect} Note that the resulting model equations of our APFC model are not rotationally
       invariant -- just like the ones of the famous WBK model \cite{wheeler}.
       This might cause problems when studying for example several grains
       and issues of their orientation with respect to each other might 
       play an important role.
\end{thebibliography}
\end{document}